\documentclass[3p,times,procedia]{elsarticle}
\usepackage{amsmath}
\usepackage{graphicx}
\usepackage{subfigure} 
\usepackage[figuresright]{rotating}

\begin{document}
\begin{frontmatter}

\title{A Pedagogical Discussion on Neutrino Wave-Packet Evolution}
\author{Cheng-Hsien Li\corref{cor1}}
\ead{cli@physics.umn.edu}
\cortext[cor1]{Corresponding author. Tel.: +1-612-624-2327}
\author{Yong-Zhong Qian}
\ead{qian@physics.umn.edu}

\address{School of Physics and Astronomy, University of Minnesota, Minneapolis, Minnesota 55455, USA}

\begin{abstract}
We present a pedagogical discussion on the time evolution of a Gaussian neutrino wave packet in free space. 
A common treatment is to keep momentum terms up to the quadratic order in the expansion of the
energy-momentum relation so that the Fourier transform can be evaluated analytically via Gaussian integrals. 
This leads to a solution representing a flat Gaussian distribution with a constant longitudinal width and 
a spreading transverse width, which suggests that special relativity would be violated if the neutrino wave 
packet were detected on its edge. However, we demonstrate that by including terms of higher order in
momentum the correct geometry of the wave packet is restored. The corrected solution has 
a spherical wave front so that it complies with special relativity.
\end{abstract}


\end{frontmatter}
\section{Introduction}
\label{sec: intro}
\begin{figure}[h!]
\begin{center}
\includegraphics[scale=0.28]{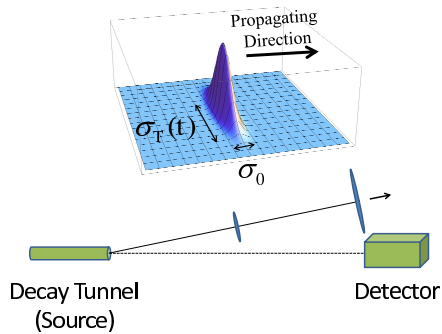}
\caption{\label{fig:Off-Axis Neutrino Detected}An off-axis neutrino WP may be detected due to 
its spreading in the transverse direction.}
\end{center}
\end{figure}

After the OPERA Collaboration reported their false superluminal neutrino results in 2011, 
the authors of \cite{Naumov:2011mm} claimed that the off-axis neutrinos may be detected 
earlier than those traveling along the source-detector axis due to the spreading of a neutrino 
wave packet (WP) in the transverse direction. For a WP with a finite size, its momentum 
uncertainty allows it to expand transversely as it propagates in space. 
These authors claimed that, when the edge of 
an off-axis WP reaches the detector, the distance traveled by the center of the WP is shorter 
than the source-detector distance, and therefore, a superluminal effect is observed as depicted 
in Fig.~\ref{fig:Off-Axis Neutrino Detected}. 

\pagebreak
The argument was soon questioned by the author of \cite{Bernard:2011qi}, who pointed out that 
a Gaussian neutrino WP will asymptotically develop a spherical wave front. Consequently, 
the arrival time of an off-axis neutrino WP will be the same as one traveling along the 
source-detector axis. However, the argument of \cite{Bernard:2011qi} was based on the description
of a Gaussian beam that is localized only in the transverse direction but not in the longitudinal direction. 
The spherical wave front  was inferred from the phase term in the Gaussian beam solution,
which is well known in laser optics.

The evolution of an initial WP can be found by two different but equivalent methods. 
The first method involves solving a wave equation suitable to the context. 
The massless Klein-Gordon equation is appropriate in the case of neutrinos or photons propagating 
in free space. The solution usually requires some approximation to this equation and employs an 
ansatz with desired boundary conditions (e.g., \cite{PhysRevE.58.1086,563385}). The second method 
involves evolving each Fourier mode of the initial WP with a specific angular frequency and 
superposing each mode at a later time 
(e.g., \cite{Am.J.Phys.52.921,Am.J.Phys.76.1102,Christov1985364,PhysRevD.70.053010}). 
Tackling an oscillatory integral is the major difficulty of the second method.  

For pedagogical purpose, we revisit the time evolution of a Gaussian WP with a sharply peaked 
momentum distribution. In particular, we review the necessary computational techniques of the 
second method and extend the solution commonly found in literature by including higher-order 
terms in the expansion of the energy-momentum relation. We demonstrate explicitly that 
these terms lead to the spherical shape of the WP. We end with a brief discussion of 
the corrected solution.

\section{Time Evolution of a Gaussian Neutrino WP}
In order to simplify the calculations, we assume an initial neutrino WP with a momentum 
distribution sharply peaked around the average momentum. Specifically, at time $t=0$, the
neutrino state is an isotropic Gaussian WP with a spatial width $\sigma_0$ corresponding to
a momentum width $(2\sigma_0)^{-1}$. Its average momentum is
$\vec{k}_0=k_0\hat{z}$ with $k_0\sigma_0 \gg 1$. Under the above assumption,
the spinor part of the neutrino wave function is essentially a constant factor and 
can be neglected, and only the evolution of the scalar part needs to be followed.
In addition, we assume that the neutrino is massless and without flavor oscillations. 
The calculations for massive neutrinos are straightforward but introduce unnecessary 
complications to the discussion. Natural units are used throughout the presentation.

The position-space wave function for the initial neutrino WP is
\begin{eqnarray}
\label{eq: Initial state in position space}
\Psi(\vec{r},0)=\frac{1}{(2\pi\sigma_0^2)^{3/4}}
\exp\left(
-\frac{r^2}{4\sigma_0^2}+ik_0z
\right)\text{,}
\end{eqnarray}
and its Fourier transform is 
\begin{eqnarray}
\label{eq: Initial state in momentum space}
\tilde{\Psi}(\vec{k})=(8\pi\sigma_0^2)^{3/4}
\exp\left[
-\sigma_0^2 k_\perp^2 - \sigma_0^2(k_z-k_0)^2
\right]\text{,}
\end{eqnarray}
where $k_\perp^2\equiv k_x^2+k_y^2$. At a later time, each plane-wave component of the WP 
evolves with a specific angular frequency and gains a phase factor $\exp(-i\omega t)$, 
where $\omega=|\vec{k}|$ is the dispersion relation (the energy-momentum relation for a
massless neutrino). Superposing each plane-wave component 
at $t>0$, we can express the resulting wave function as
\begin{equation}
\label{eq: Wave function in position space - General form}
\Psi(\vec{r},t)=\int\frac{\mathrm{d}^3\vec{k}'}{(2\pi)^3}
\, (8\pi\sigma_0^2)^{3/4}
\exp\left[
-\sigma_0^2 k'^2-i\omega_{\vec{k'}+\vec{k}_0}t+i(\vec{k'}+\vec{k}_0)\cdot\vec{r}
\right]\text{,}
\end{equation}
where $\vec{k}'=\vec{k}-\vec{k}_0$. To simplify the notation, we will use 
$\vec{k}$ for $\vec{k}'$ below. Noting that the integrand in the above equation
is significant only for $|\vec{k}|\lesssim \mathcal{O}(\sigma_0^{-1})\ll k_0$, 
we expand the dispersion relation around $\vec{k}=0$:
\begin{equation}
\label{eq: dispersion relation expansion}
\omega_{\vec{k}+\vec{k}_0}\equiv\omega_q+\delta\omega=
k_0 \left(
1+\frac{k_z}{k_0}+\frac{k_\perp^2}{2k_0^2}
-\frac{k_zk_\perp^2}{2k_0^3}+\frac{4k_z^2k_\perp^2-k_\perp^4}{8k_0^4}+\cdots
\right)\text{,}
\end{equation}
where $\omega_q$ represents the terms up to the quadratic order in the parentheses
and $\delta\omega$ those beyond. 

If we approximate $\omega_{\vec{k}+\vec{k}_0}$ as $\omega_q$, the exponent in 
Eq.~\eqref{eq: Wave function in position space - General form} can be expressed 
as quadratic polynomials of $k_x, k_y,$ and $k_z$. Upon completing the square 
for each momentum variable, we can evaluate 
Eq.~\eqref{eq: Wave function in position space - General form} 
via Gaussian integrals to obtain the zeroth-order solution 
\begin{equation}
\label{eq: zeroth order solution}
\Psi^{(0)}(\vec{r},t)=
\frac{1}{(2\pi\sigma_0^2)^{3/4}\left[1+it(2k_0\sigma_0^2)^{-1}\right]}
\exp\left\{
-\frac{x^2+y^2}{4\sigma_0^2\left[1+it(2k_0\sigma_0^2)^{-1}\right]}
-\frac{(z-t)^2}{4\sigma_0^2}+ik_0(z-t)
\right\}\text{.}
\end{equation}  
This solution is a plane wave with momentum $\vec{k}_0$ modulated by a complex 
envelope that is traveling along the $z$-direction at the speed of light. 
The modulus squared of Eq.~\eqref{eq: zeroth order solution}, or the probability density, 
essentially corresponds to a ``flat disc'' with a time-dependent transverse width 
$\sigma_T(t)=\sigma_0\sqrt{1+t^2(4k_0^2\sigma_0^4)^{-1}}$ as shown in 
Fig.~\ref{fig:Off-Axis Neutrino Detected}.   

To go beyond the zeroth-order solution, we include $\delta\omega$ in 
$\omega_{\vec{k}+\vec{k}_0}$ and use
$\exp(i\delta\omega t)=\sum_{n=0}^\infty (i\delta\omega t)^n/n!$ to rewrite
Eq.~\eqref{eq: Wave function in position space - General form} as
\begin{equation}
\label{eq: Wave function in position space - with correction}
\begin{split}
\Psi(\vec{r},t)=\int\frac{\mathrm{d}^3\vec{k}}{(2\pi)^3}
\, (8\pi\sigma_0^2)^{3/4}&
\exp\left\{
-\sigma_0^2 k^2-i\omega_{q}t+i(\vec{k}+\vec{k}_0)\cdot\vec{r}
\right\} \\
&\times\left[1+\frac{it}{2k_0^2}k_zk_\perp^2
-\frac{it}{8k_0^3}\left(4k_z^2k_\perp^2-k_\perp^4\right)
+\text{higher orders}\right]
\text{,}
\end{split}
\end{equation}
which again can be evaluated via Gaussian integrals. 
The correction to $\Psi^{(0)}(\vec{r},t)$ can be efficiently computed by using 
the following identity:
\begin{eqnarray}
\label{eq: ratio of Gaussian integrals}
\frac{\int_{-\infty}^\infty\mathrm{d}k\,k^n
\mathrm{e}^{-a(k-b)^2+c}}{\int_{-\infty}^\infty\mathrm{d}k\,\mathrm{e}^{-a(k-b)^2+c}}
=\sum_{l=0}^{n}\binom{n}{l}b^{n-l}\times
\begin{cases}
0,&\text{for odd}\ l,\\
(l-1)!!/(2a)^{l/2},&\text{for even}\ l.
\end{cases}
\end{eqnarray} 
The complex coefficients $a$, $b$, and $c$ in the above equation are obtained 
from completing the square for each momentum variable in the exponent of
Eq.~\eqref{eq: Wave function in position space - with correction}.
In principle, $\Psi(\vec{r},t)$ can be calculated to any desired order:
$\Psi(\vec{r},t)=\Psi^{(0)}(\vec{r},t)+\Psi^{(1)}(\vec{r},t)+\cdots$.
For illustration, we present here only the leading-order correction:
\begin{equation}
\label{eq: corrected solution}
\Psi^{(1)}(\vec{r},t)=-\frac{t}{4k_0^2\sigma_0^3}\times
\frac{(z-t)}{\sigma_0\left[1+it(2k_0\sigma_0^2)^{-1}\right]}
\left\{
1
-\frac{x^2+y^2}{4\sigma_0^2\left[1+it(2k_0\sigma_0^2)^{-1}\right]}
\right\}\Psi^{(0)}(\vec{r},t)\text{,}
\end{equation}
which is obtained from the second term in the square brackets of 
Eq.~\eqref{eq: Wave function in position space - with correction}.

\begin{figure}[h!]
\begin{center}
\subfigure[]{
\includegraphics[width=0.40\columnwidth]{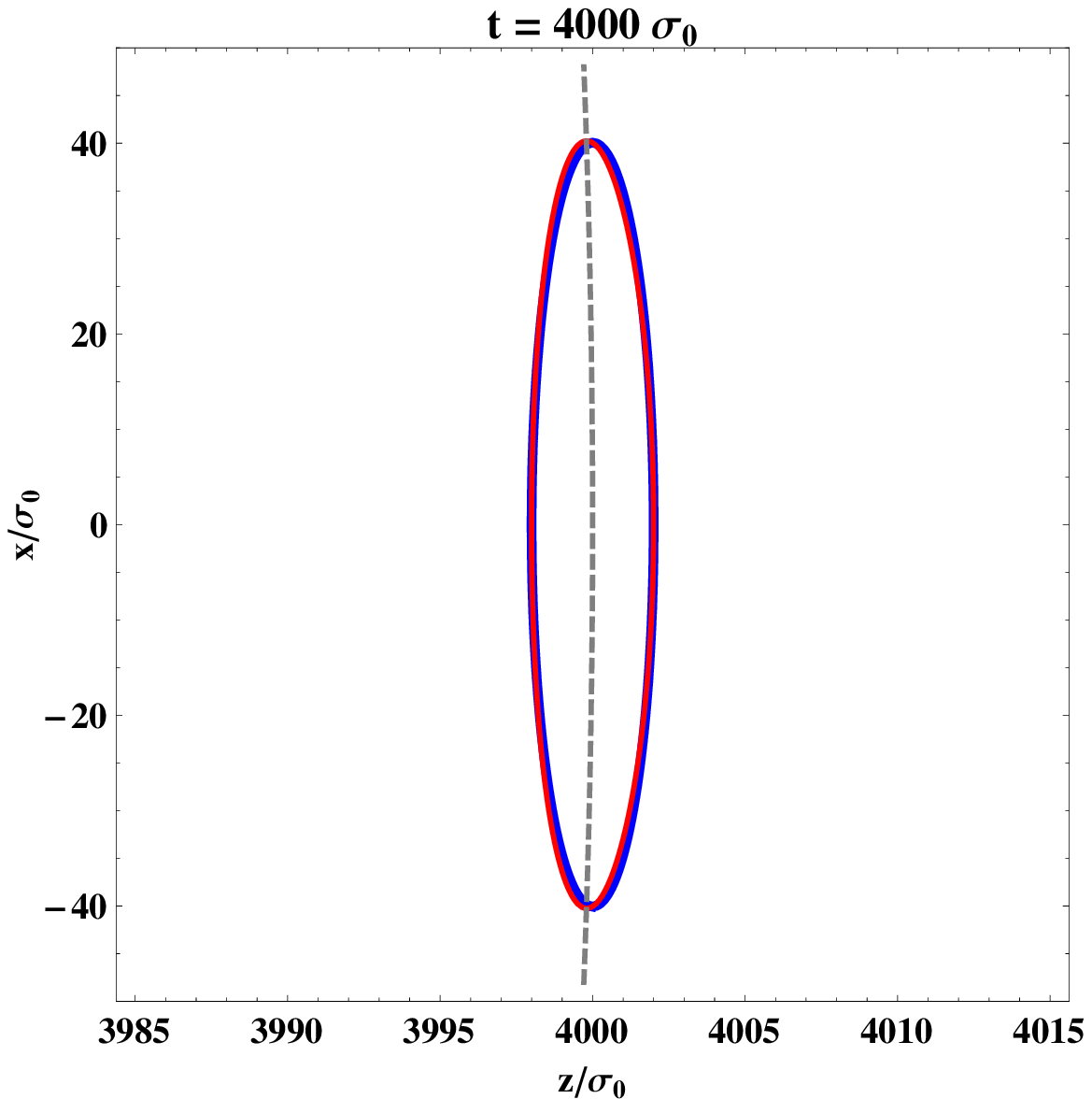}}
\subfigure[]{
\includegraphics[width=0.405\columnwidth]{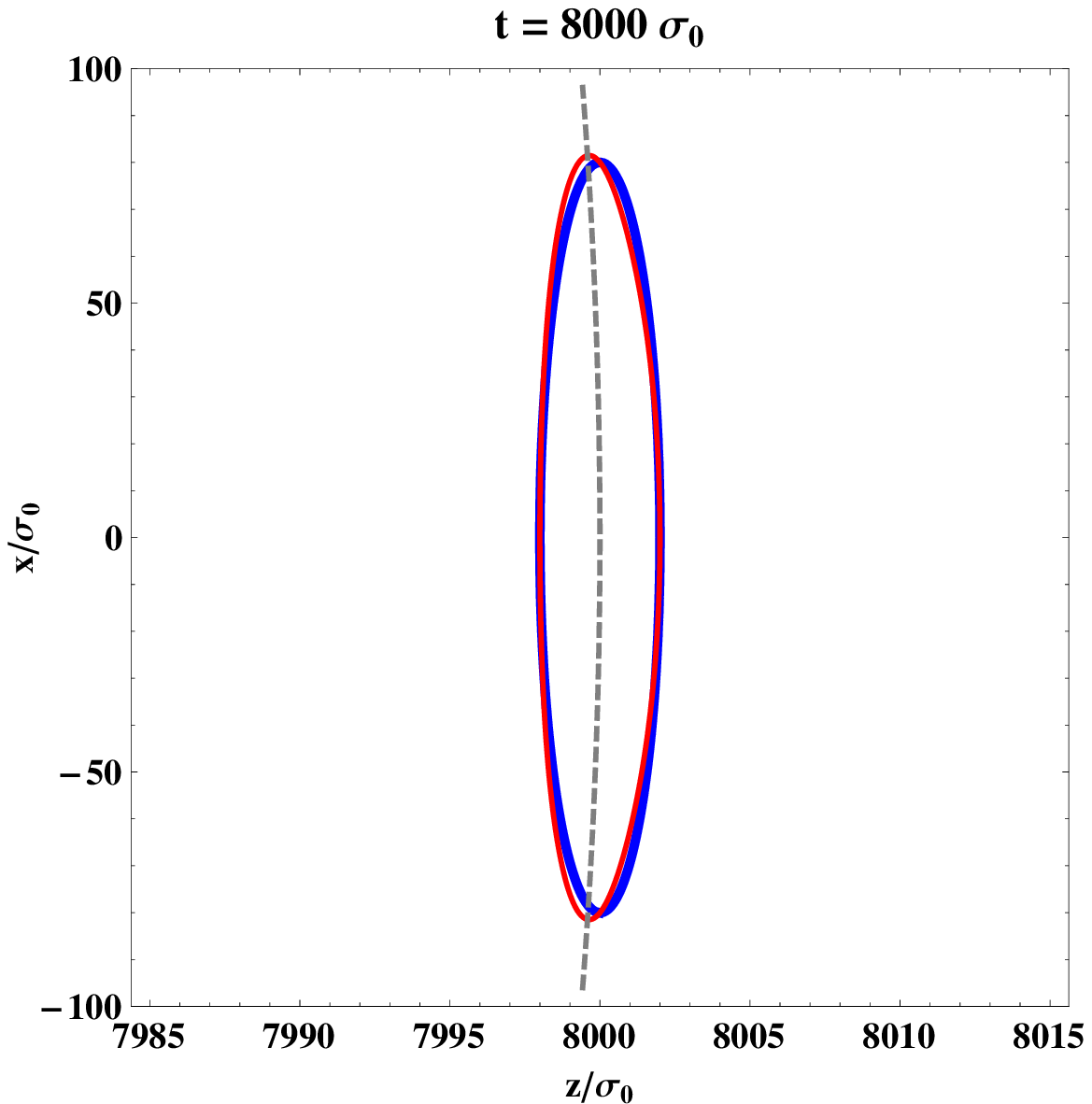}}
\subfigure[]{
\includegraphics[width=0.40\columnwidth]{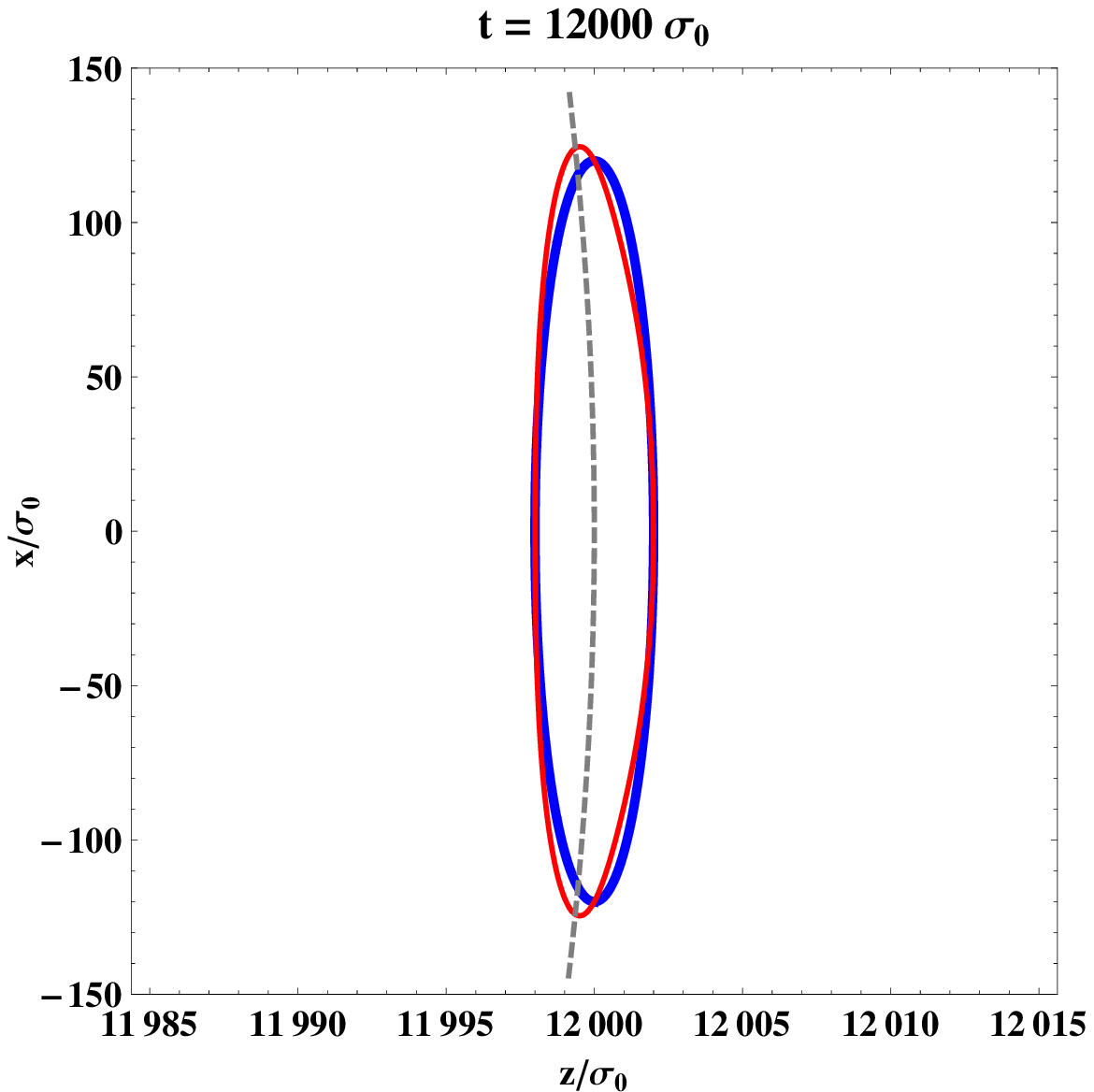}}
\subfigure[]{
\includegraphics[width=0.40\columnwidth]{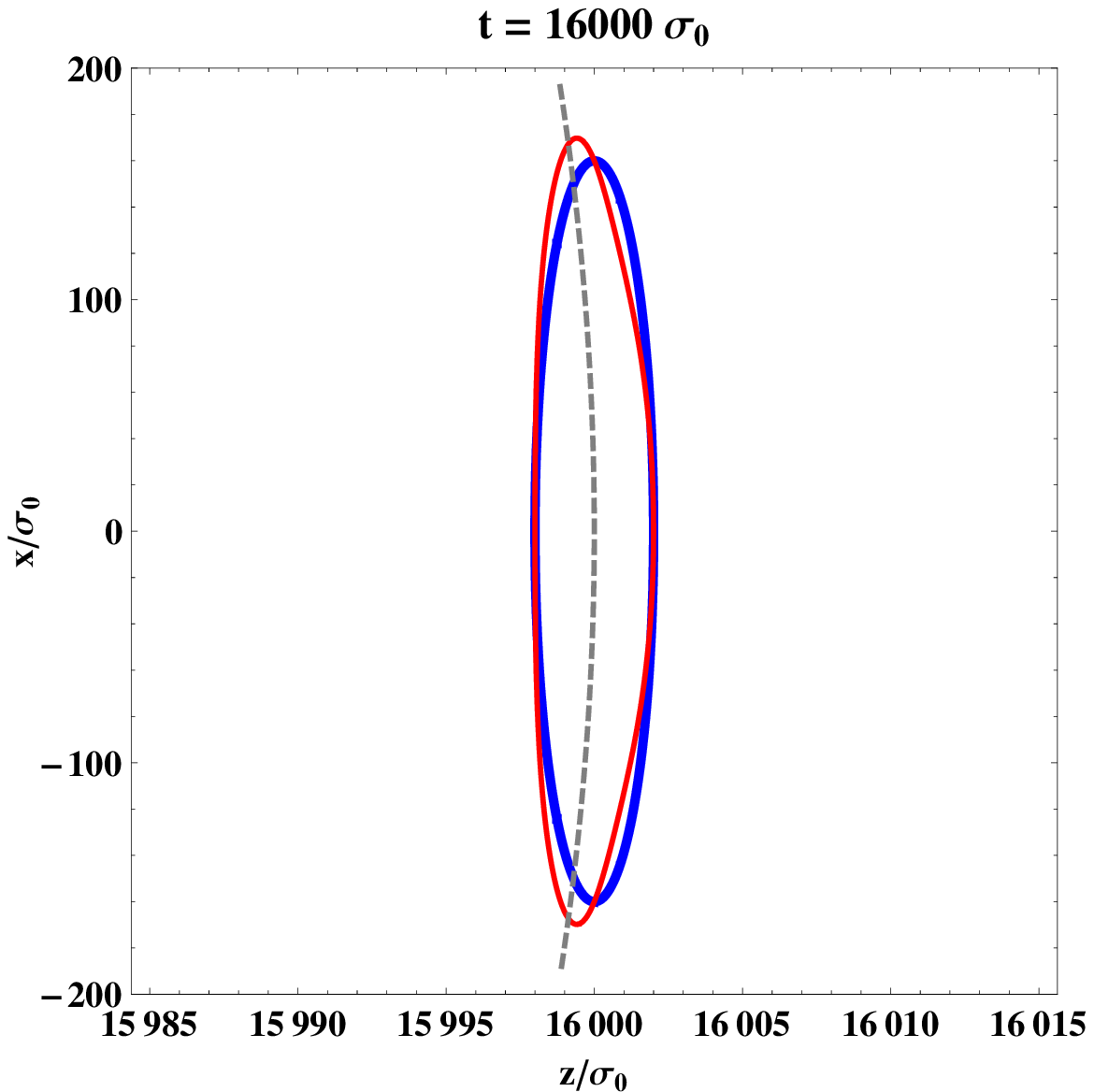}}
\caption{\label{fig:Time Evolution of WP}
The $2\sigma$ contours of the probability density calculated from 
$\Psi^{(0)}(\vec{r},t)$ and $\Psi(\vec{r},t)$ are shown in
blue and red, respectively. 
The gray dashed curve indicates a circular arc of radius $t$. 
These plots are for $y=0$ and assume $k_0\sigma_0=100$. 
The horizontal scale is the same for all figures while the vertical 
scale is set in proportion to the increasing transverse size of the WP.}
\end{center}
\end{figure}

\section{Discussion}
A comparison of the contour of constant probability density for 
$\Psi^{(0)}(\vec{r},t)$ and $\Psi(\vec{r},t)$ is made in Fig.~\ref{fig:Time Evolution of WP}. 
We can observe that significant deviation of $\Psi(\vec{r},t)$ from $\Psi^{(0)}(\vec{r},t)$ 
appears when the phase correction $\delta\omega t$ in 
Eq.~\eqref{eq: Wave function in position space - with correction} is no longer negligible. 
The leading correction term in Eq.~\eqref{eq: corrected solution} is proportional to $(z-t)$, 
which gives rise to the asymmetry between the space-like and time-like regions, thereby
``bending'' the flat Gaussian distribution. Thus the initial Gaussian WP develops a spherical 
wave front after higher-order terms in the phase are taken into account.

The result from Eq.~\eqref{eq: Wave function in position space - with correction}
is ensured by the infinitely large convergence radius of the complex exponential function.
It is just a matter of determining how many correction terms are required in order to 
produce a good approximation. To address this issue, we define $\delta\omega^{(i)}$
to correspond to the $i{\text{th}}$-order term ($i\geq 3$) in the parentheses of 
Eq.~\eqref{eq: dispersion relation expansion}.
The result in Eq.~\eqref{eq: corrected solution} obtained with 
the approximation $\exp(i\delta\omega t)\approx 1+i\delta\omega^{(3)}t$
is valid for $|\delta\omega^{(3)}t|< 1$. This implies a limit on the propagation time  
$t\lesssim2k_0^2\sigma_0^3$. It is not difficult to envision that, as time increases, 
a tiny contribution to $\delta\omega$ could totally change the value of 
$\exp(i\delta\omega t)$. Unfortunately, if the propagation time exceeds 
$2k_0^2\sigma_0^3$, the required computational effort to obtain $\Psi(\vec{r},t)$
from Eq.~\eqref{eq: Wave function in position space - with correction} dramatically increases.

If one would like to find an acceptable solution at $t>2k_0^2\sigma_0^3$ through 
the laborious computation using 
Eqs.~\eqref{eq: Wave function in position space - with correction} and 
\eqref{eq: ratio of Gaussian integrals}, a reasonable procedure is as follows:
\begin{enumerate}
\item Find the term $\delta\omega^{(m)}$ in the expansion series of 
$\omega_{\vec{k}+\vec{k}_0}$
such that $\pi\sim|\delta\omega^{(m)}t|\gg |\delta\omega^{(m+1)}t|$
based on the estimate
$|\delta\omega^{(m)}t|\sim\mathcal{O}\left(k_0 t(k_0\sigma_0)^{-m}\right)$. 
Make the approximation $\exp(i\delta\omega t)\approx \exp(i\delta\tilde{\omega} t)$,
where $\delta\tilde{\omega}
\equiv\sum_{i=3}^m\delta\omega^{(i)}$.
\item Make another approximation
$\exp(i\delta\tilde{\omega} t)\approx\sum_{n=0}^p (i\delta\tilde{\omega} t)^n/n!$,
where the integer $p$ should be at least an order of magnitude larger than 
$|\delta\omega^{(3)}t|$ to ensure good convergence.
\item Compute the contribution to $\Psi(\vec{r},t)$ in 
Eq.~\eqref{eq: Wave function in position space - with correction} from each term in 
$\sum_{n=0}^p (i\delta\tilde{\omega} t)^n/n!$ using Eq.~\eqref{eq: ratio of Gaussian integrals}.
\end{enumerate} 
We have not found an efficient way to simplify the cumbersome algebra in the above steps
and conclude that the perturbative method discussed above is practical only for limited propagation time.

By evolving the different Fourier modes in an initial Gaussian neutrino WP,
we have shown that the WP develops a spherical wave front, thereby obeying
the limit on its propagation speed imposed by special relativity. We note that
the apparent overlap of the WP with the space-like region shown in 
Fig.~\ref{fig:Time Evolution of WP} must not be mistaken as a superluminal effect:
this is purely due to the intrinsic position uncertainty of the initial WP.
We also note that the transverse expansion of the WP indeed allows detection of 
the neutrino over an increasing spatial extent. This may have interesting
consequences for long-baseline neutrino experiments, which we plan to
study in the future.

\section*{Acknowledgments} 
This work was supported in part by the US DOE Award No. DE-FG02-87ER40328.

\bibliographystyle{elsarticle-num}
\bibliography{ref}

\end{document}